\documentclass[letterpaper, 10 pt, conference]{ieeeconf}  % Comment this line out
\usepackage{amssymb,color}
\usepackage{cite,paralist}
\usepackage{graphicx,mathrsfs}
\usepackage{subfigure}
\usepackage{mathrsfs}
\usepackage{amsmath}
\usepackage{amsfonts}
\usepackage{subfigure}

\newtheorem{theorem}{{Theorem}}
\newtheorem{lemma}[theorem]{{Lemma}}

\newcommand{\mb}{\mathbf}
\newcommand{\qed}{\hspace*{\fill} $\Box$ \\}

\newcommand{\nix}[1]{}
\allowdisplaybreaks

\title{\LARGE \bf
Network Coding Capacity of Random Wireless Networks under a Signal-to-Interference-and-Noise
Model}

\author{\authorblockN{Zhenning Kong\authorrefmark{1},
Salah A. Aly\authorrefmark{2}, Emina Soljanin\authorrefmark{3},
Edmund M. Yeh\authorrefmark{1},  and Andreas
Klappenecker\authorrefmark{2} }
\authorblockA{\authorrefmark{1}Department of Electrical Engineering, Yale University,
New Haven, CT 06520, USA \\
Email: \{zhenning.kong, edmund.yeh\}@yale.edu}
\authorblockA{\authorrefmark{2}Department of Computer Science, Texas A\&M University,
College Station, TX 77843, USA \\
Email: \{salah, klappi\}@cs.tamu.edu}
\authorblockA{\authorrefmark{3}Bell Laboratories, Alcatel-Lucent,
Murray Hill, NJ 07974, USA \\
Email: emina@lucent.com}}

\begin{document}

\maketitle
\thispagestyle{empty}
\pagestyle{empty}

\begin{abstract}
In this paper, we study network coding capacity for random wireless
networks. Previous work on network coding capacity for wired and
wireless networks have focused on the case where the capacities of
links in the network are independent. In this paper, we consider a
more realistic model, where wireless networks are modeled by random
geometric graphs with interference and noise. In this model, the
capacities of links are not independent.  We consider two scenarios,
single source multiple destinations and multiple sources multiple
destinations. In the first scenario,  employing coupling and
martingale methods, we show that the network coding capacity for
random wireless networks still exhibits a concentration behavior
around the mean value of the minimum cut under some mild conditions.
Furthermore, we establish upper and lower bounds on the network
coding capacity for dependent and independent nodes. In the second
one, we also show that the network coding capacity still follows a
concentration behavior. Our simulation results confirm our
theoretical predictions.
\end{abstract}

\section{Introduction}

Network coding was originally proposed by Ahlswede~\emph{et al.} in~\cite{ahlswede00}. Unlike
traditional store-and-forward routing algorithms, in network coding schemes, intermediate nodes
encode their received messages and forward the coded messages to their next-hop neighbors. It has
been shown that network coding can improve the network capacity, even by using simple linear or
random codes~\cite{LiYeCa03, KoMe03, JaSaChEfEgTo05, HoKoMeEfShKa06}. In most studies of network
coding, network topologies are assumed to be known.

In~\cite{RaShWe03, RaShWe05}, the authors studied network coding capacity for weighted random
graphs and random geometric graphs. In the random graph model, each pair of nodes are connected by
a bidirectional link with probability $p<1$ independently~\cite{Bo01,JaLuRu00}. The capacity of
each link is assumed to be i.i.d. according to some probability distribution. In the random
geometric graph model, two nodes are connected to each other by a bidirectional link only when
their distance is less than a predefined positive value $r$, the characteristic
radius~\cite{Pe03}. Each link has a unit capacity. For these two types of random networks, the
authors showed that the network coding capacity is concentrated at the (weighted) mean degree of
the graph, i.e., the (weighted) mean number of neighbors of each node. Essentially, the results
reveal a concentration behavior of the size of the minimum cut between two nodes in random graphs
or random geometric graphs. Similar problems have been studied in the literature, e.g.,
\cite{DiPePeSe01} and references there. In~\cite{AlKaMeKl07}, the authors studied a generalized
random geometric graph model, where two nodes are connected by a bidirectional link with
probability 1 if their distance $d$ is less than $r_0>0$ and with probability $p<1$ if $r_0<d\leq
r_1$. They obtained similar concentration results there.

The geometric models in~\cite{RaShWe03, RaShWe05, AlKaMeKl07} assume that a link exists (possibly
with a probability) between two nodes when the nodes are within each other's transmission range.
Although each link has a direction, as all links are bidirectional (i.e., the link $(i,j)$ implies
the existence of the link $(j,i)$), the model in fact leads to an {\em undirected} graph and
considerably simplifies the resulting analysis. In addition, interferences among wireless
terminals were not considered in~\cite{RaShWe03, RaShWe05, AlKaMeKl07}. Nevertheless, in wireless
networks, due to noise, interference, and heterogeneity of transmission power, significantly more
sophisticated models for link connectivity are needed. For instance, a widely-used model for
wireless communication channels is the Signal-to-Interference-plus-Noise-Ratio (SINR)
model~\cite{Pr00, TsVi05}. In this paper, we study the capacity, i.e., the size of the minimum
cut, of random wireless networks under the SINR model.

Since how to apply the network coding with noisy links is still an open problem, we assume that as
long as the SINR of a link $(i,j)$, $\beta_{ij}$ is greater than or equal to a predefined
threshold $\beta$, then node $i$ can transmit data at rate $R$ packets/sec to node $j$ without any
error. That is links are noise-free once the SINR condition is met. In other words, we view the
network coding as operation on a higher layer in the network communication stack, and assume there
is an error correcting code at the lower layer which corrects errors on the links once the SINR
threshold is met. Then, in this model, each link is indeed directional (not necessarily
bidirectional), and the capacities of different links are not independent. We will show that the
capacity still has a sharp concentration when the scale of the network is large enough.

This paper is organized as follows. In Section II, we describe the
random wireless network model. In Section III, we study the network
coding capacity for a single source and multiple destinations
transmissions. Specifically, we investigate two cases. In the first
one, all nodes have the same transmission power, and in the second
one, the transmission powers are heterogeneous. We use different
techniques for these two cases and show that the network coding
capacity has a concentration behavior in both cases. In Section IV,
we extend our result to multiple sources and multiple destinations
transmission problem. In Section V, we present some simulation
results, and finally, we conclude this paper in Section VI.

\section{Random Wireless Networks Model}

We use the following model for random wireless networks. Assume
\begin{itemize}
\item[(i)] $\mathcal{X}=\{{\mb X}_1, {\mb X}_2, ..., {\mb X}_n\}$ is a set of i.i.d.
two-dimensional random variables according to a homogeneous Poisson point process in the
two-dimensional unit torus, where ${\mb X}_i$ denotes the random location of node $i$, and $n$ is the total number of nodes.
\item[(ii)]
Each node $i$ has a transmission power $P_i$, which follows a probability distribution $f_P(p)$,
$p\in [p_{min},p_{max}]$, where $0<p_{min}\leq p_{max}<\infty$.
\end{itemize}
Here, the existence of a link from node $i$ to node $j$ depends on the ability to decode the
transmitted signal from $i$ to $j$, which is determined by the Signal-to-interference-plus-noise-ratio (SINR) given by
\begin{equation}\label{eq:beta-ij}
\beta_{ij}=\frac{P_iL(d_{ij})}{N_0+\gamma \sum_{k\neq i,j} P_kL(d_{kj})},
\end{equation}
where $P_i$ is the transmission power of node $i$, $d_{ij}$ is the distance between nodes $i$ and
$j$, and $N_0$ is the power of background noise. The parameter $\gamma$ is the inverse of system
processing gain.  It is equal to 1 in a narrow-band system and smaller than 1 in a broadband
(e.g., CDMA) system. The signal attenuation function $L(\cdot)$ is a function of the distance
$d_{ij}=||\mathbf{X}_i-\mathbf{X}_j||$, where $\|\cdot\|$ is the Euclidean norm, and is usually
given by $L(d_{ij}) = cd_{ij}^{-\alpha}$ for some constants $c$ and $2 < \alpha < 4$.

Under the SINR model, the transmitted signal of node $i$ can be decoded at $j$ if and only if
$\beta_{ij} > \beta$, where $\beta$ is some threshold for decoding. In this case, a link $(i,j)$
is said to exist from $i$ to $j$. Note that even if $\beta_{ij} > \beta$, $\beta_{ji} > \beta$ may
not hold and thus the link $(j,i)$ may not exist. Thus, the graph resulting from the SINR model is
in general {\em directed.} It is clear that link $(i,j)$ is bidirectional if and only if
$\min\{\beta_{ij}, \beta_{ji}\}>\beta$. Denote by $G(\mathcal{X},\mathcal{P},\gamma)$ the ensemble
of random wireless networks induced by the above physical model, where
$\mathcal{P}=\{P_1,P_2,...,P_n\}$ represents the set of transmission power.

For transmission power $P$ and signal attenuation function $L(\cdot)$, we assume
\begin{itemize}
\item[(i)] $p_{min}>\beta N_0$; \item[(ii)] $\Pr(P=p_{min})>0, \Pr(P=p_{max})>0$, \item[(iii)]
$L(x)$ is continuous and strictly decreasing in $x$
\end{itemize}
for technical and practical reasons. In the remainder of this paper, under different
circumstances, we may add further constrains on $P_i$.

The sum $\sum_{k\neq j}L(d_{kj})=\sum_{k\neq j}L(||\mathbf{X}_k-\mathbf{X}_j||)$ is a random
variable depending on the locations of all nodes in the network. Define
\begin{equation}\label{eq:J-j}
J(j)\triangleq\sum_{k\neq j}L(d_{kj}), \quad \mbox{
for all } j,
\end{equation}
\begin{equation}\label{eq:I-j}
I(j)\triangleq\sum_{k\neq j}P_kL(d_{kj}), \quad
\mbox{ for all } j.
\end{equation}

To study the asymptotic network capacity, we will let the number of nodes $n$ go to infinity.
Since the region is fixed, this corresponds to a dense network model~\cite{Pe03, GuKu00}. Another
widely used model is the extended network model~\cite{MeRo96, DoFrTh05}, in which the number of
nodes and the area of the region both go to infinity while the ratio between them---the density of
the network, is kept as a constant. Both models are widely used in the literature. We will focus
on the former one in this paper.

\section{Network Coding Capacity for Single Source Transmission}

\subsection{Capacity of a Cut}

Let $C_{ij}$ be the capacity of a link $(i,j)$. We will specify the value of $C_{ij}$ later for
different scenarios. Consider a single-source multiple-destination transmission problem. Let $s$
be the source node. Suppose there are $l$ destination nodes, $t_1,...t_l$, and $m$ relay nodes,
$u_1,...u_m$. Denote the set of the destination nodes and relay nodes by $\mathcal{T}$ and
$\mathcal{R}$, respectively. Fig.~1 illustrates an example of single-source single-destination
transmission.

\begin{figure}[t]
\centering
\includegraphics[width=2.5in]{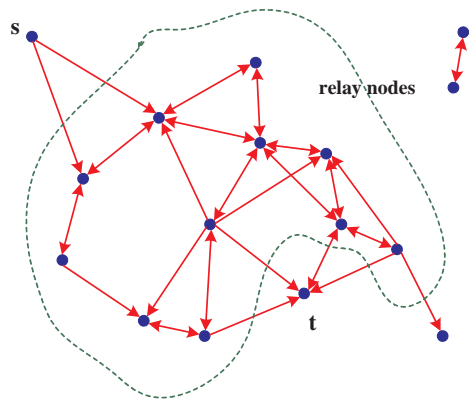}
\caption{Single-source single-destination transmission in directed SINR graphs}
\end{figure}

Let the capacity of the link from the source $s$ to each relay node $u_i$ be $C_{si}$,
$i=1,...,m$, the capacity from relay node $u_i$ to another relay node $u_j$ be $C_{ij}$, $i\neq j,
i=1,...,m, j=1,...,m$, and the capacity from each relay node $u_i$ to each destination node $t_j$
be $C_{it_j}$, $i=1,...,m, j=1,...,l$. Unlike random geometric graph models studied
in~\cite{RaShWe03,RaShWe05, AlKaMeKl07}, the capacities in our model are not symmetric nor
independent in general.

Since in our random SINR wireless network model, there are two sources of randomness: one is the
random location of each node and the other is the random transmission power of each node. We use
$E_{X}$ and $E_{P}$ to denote the expectation operation with respect to each probability measure
respectively.

Let $\bar{C}$ be the expected capacity of a link $(i,j)$ which is defined as
\begin{eqnarray}\label{C-bar}
\bar{C} & = & E_{X}E_P[C_{ij}]\nonumber \\
& = & \int_0^{\frac{p_{max}}{N_0}}C_{ij}dF_{\beta_{ij}}(\tau),
\end{eqnarray}
where $F_{\beta_{ij}}(\cdot)$ is the c.d.f. of $\beta_{ij}$, which is determined by $f_P(\cdot)$,
the distribution of $\mathcal{X}$, and path-loss function $L(\cdot)$.

Now define an $s$-$t$-cut of size $k$ for a pair of given source $s$ and destination $t\in
\mathcal{T}$ as a partition of the relay nodes into two sets $V_k$ and $V_k^c$, such that
$|V_k|=k,|V_k^c|=m-k$, $V_k \cup V_k^c=\mathcal{R}$ and $V_k\cap V_k^c=\emptyset$. An example of
an $s$-$t$-cut is shown in Fig.~2. Let
\begin{equation}\label{eq:C-k}
C_k=\sum_{u_i\in V_k^c}C_{si}+\sum_{u_j\in V_k}\sum_{u_i\in V_k^c}C_{ji}+\sum_{u_j\in V_k}C_{jt},
\end{equation}
then $C_k$ is the capacity of the corresponding $s$-$t$-cut. Although $C_k$ is a sum of dependent random variables, we still have
\begin{eqnarray}\label{eq:C-k-expectation}
E[C_k] & = & E_{X}E_P[C_k]\nonumber\\
& = & \sum_{u_i\in V_k^c}E_{X}E_P[C_{si}]+\sum_{u_j\in V_k}\sum_{u_i\in
V_k^c}E_{X}E_P[C_{ji}]\nonumber\\
& &+\sum_{u_j\in V_k}E_{X}E_P[C_{jt}]\nonumber\\
& = & [m+k(m-k)]\bar{C},
\end{eqnarray}
and consequently $E[C_k]=E[C_{m-k}]$ for $k=0,1,...,m$, and $E[C_0]\leq E[C_1]\leq \cdots \leq
E[C_{\lceil m/2 \rceil}]$.

\begin{figure}[t]
\centering
\includegraphics[width=2.5in]{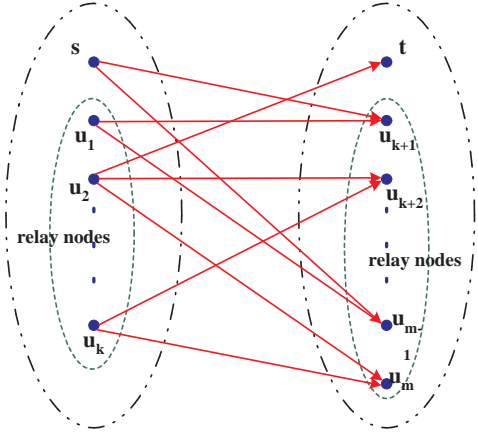}
\caption{An $s$-$t$-cut for the single-source single-destination transmission in directed SINR
graphs}
\end{figure}

To show the capacity of any source-destination pair concentrates at some value, we will first show
that for such a source-destination pair, the capacity of any $s$-$t$-cut of size $k$ concentrates
at its mean value. Similar results were proved in~\cite{RaShWe03, RaShWe05, AlKaMeKl07}, where the
capacities of the links that originate from the same node are i.i.d. Nevertheless, the methods
used in~\cite{RaShWe03, RaShWe05, AlKaMeKl07} do not apply here, since in the SINR model, $C_k$ is
a sum of dependent random link capacities. Instead, we employ coupling, martingale methods and
Azuma's inequality~\cite{Pe03, MoRa95} to solve the problem for different cases.

Note that when $\gamma=0$, i.e., there is no interference in the networks, the capacities $C_{si}$
for $i=k+1,...,m$ are mutually independent, as well as the capacities $C_{ij}$ for any fixed
$i=1,...,k$ with $j=k+1,...,m$ or $t$. In this case, although the link capacities are still
asymmetric, $\sum_{u_i\in V_k^c}C_{si}$ and $\sum_{u_i\in V_k^c\cup \{t\}}C_{ji}$ for $j\in V_k$
become sums of independent random variables. Thus we can apply methods similar to those used
in~\cite{RaShWe03,RaShWe05,AlKaMeKl07} to obtain the same concentration results.

\subsection{Constant Transmission Power}

Consider the scenario when all nodes transmit with a constant power $P_0$ and denote the model by $G(\mathcal{X},P_0,\gamma)$. The SINR of link $(i,j)$ in this case,
$\beta_{ij}$, can be rewritten as
\begin{eqnarray}\label{eq:beta-ij-Constant-Power}
\beta_{ij} & = & \frac{L(d_{ij})}{N_0/P_0+\gamma \sum_{k\neq i,j} L(d_{kj})} \nonumber \\
& = & \frac{L(d_{ij})}{N_0/ P_0+\gamma J(j)-\gamma L(d_{ij})}.
\end{eqnarray}

Assume when $\beta_{ij}\geq \beta$, the link $(i,j)$ has  a capacity $R$, i.e., node $i$ can
transmit data at rate $R$ packets/sec to node $j$ without any error. Then, we can define $C_{ij}$
as
\begin{equation}\label{eq:C-ij-R}
C_{ij}=\left\{\begin{array}{ll} R & \beta_{ij}\geq \beta,\\ 0 & \beta_{ij}<\beta.
\end{array}\right.
\end{equation}

Note that when the wireless channel is Gaussian channel, the capacity of link $(i,j)$
is~\cite{CoTh91}
\begin{equation}\label{eq:C-ij-Gaussian}
C_{ij}=\left\{\begin{array}{ll} \vspace{+.1in} \displaystyle
\frac{1}{2}\log\left(1+\beta_{ij}\right) & \beta_{ij}\geq \beta,\\ 0 & \beta_{ij}<\beta.
\end{array}\right.
\end{equation}

Our results in this subsection do not rely on any particular expression of $C_{ij}$, and thus they
hold for $C_{ij}$ defined by \eqref{eq:C-ij-R} as well as for $C_{ij}$ defined by
\eqref{eq:C-ij-Gaussian}. Nevertheless, since we consider the application of network coding, it
would be more appropriate to focus on the $R-0$ capacity \eqref{eq:C-ij-R}, rather than the
capacity of Gaussian channel.

Note that $\beta_{ij}$ and thus $C_{ij}$ are determined by $L(d_{ij})$ and $J(j)$. Because of the
i.i.d. distribution of $\mathbf{X}_i$'s, given $\mathbf{X}_j$, $d_{ij}$'s are independent for all
$i \neq j$. Given node $j$, let
\begin{equation}\label{eq:E-L}
E[L]\triangleq E_{\mathbf{X}_i}[L(d_{ij})],
\end{equation}
then
\begin{equation}\label{eq:E-J}
E[J(j)] = E\left[\sum_{i\neq j}L(d_{ij})\right]=(n-1)E[L]\triangleq E[J].
\end{equation}

Since our model is a dense network model and the area of the region is fixed, $E[L]=E[L(d_{ij})]$
is a constant and $E[J]=(n-1)E[L]$ scales with $n$. For different $j$'s, it is clear that $J(j)$'s
are not independent, however, they have the same sharp concentration behavior in large scale
wireless networks. This is established in the following lemma.

\vspace{+.1in}
\begin{lemma}\label{Lemma:J-j-bound}
Suppose there are $n$ nodes in the network, then
\begin{equation}\label{eq:J-j-concentration-lower}
\Pr(J(j)\leq (1-\epsilon_1)E[J])=O\left(\frac{1}{n^2}\right),
\end{equation}
and
\begin{equation}\label{eq:J-j-concentration-upper}
\Pr(J(j)\geq (1+\epsilon_1')E[J])=O\left(\frac{1}{n^2}\right),
\end{equation}
for all $j=1,2,...,n$, where $\epsilon_1=\sqrt{\frac{4\ln n}{(n-1)E[L]}}$ and $\epsilon_1'=\sqrt{\frac{6\ln n}{(n-1)E[L]}}$.
\end{lemma}
\vspace{+.1in}

\emph{Proof:}  Given any node $j$, because $J(j)=\sum_{i\neq j}L(d_{ij})$, and $L(d_{ij})$ are
i.i.d. for all $i\neq j$, by the Chernoff bound~\cite{MoRa95, AlSp00}, we have
\begin{eqnarray}\label{eq:J-j-Chernoff-lower-bound}
\Pr(J(j)\leq (1-\epsilon_1)E[J]) \!\!\!\! & \leq & \!\!\!\! \exp\left\{-\frac{E[J]\epsilon_1^2}{2}\right\}\nonumber\\
 \!\!\!\! & = &  \!\!\!\!\exp\left\{-\frac{(n-1)E[L]\epsilon_1^2}{2}\right\}
\end{eqnarray}
and
\begin{eqnarray}\label{eq:J-j-Chernoff-upper-bound}
\Pr(J(j)\geq (1+\epsilon_1')E[J]) \!\!\!\! & \leq &  \!\!\!\! \exp\left\{-\frac{E[J]\epsilon_1'^2}{3}\right\}\nonumber\\
\!\!\!\! & = & \!\!\!\! \exp\left\{-\frac{(n-1)E[L]\epsilon_1'^2}{3}\right\}.
\end{eqnarray}

Substituting $\epsilon_1=\sqrt{\frac{4\ln n}{(n-1)E[L]}}$ and $\epsilon_1'=\sqrt{\frac{6\ln
n}{(n-1)E[L]}}$ into \eqref{eq:J-j-Chernoff-lower-bound} and \eqref{eq:J-j-Chernoff-upper-bound},
we obtain \eqref{eq:J-j-concentration-lower} and \eqref{eq:J-j-concentration-upper}, respectively. \qed

Lemma \ref{Lemma:J-j-bound} shows that when the network is large, i.e., $n$ is sufficiently large,
the interference at each node concentrates at $\gamma (n-1)E[L]=\Theta(1) E[L]$. The reason for
this is the uniformly (asymptotically Poisson) random distribution of the nodes.

Now define two other types of SINR models $G'(\mathcal{X},P_0,\gamma)$ and
$G''(\mathcal{X},P_0,\gamma)$ which are coupled with $G(\mathcal{X},P_0,\gamma)$ such that they
have the same point process $\mathcal{X}$ and constant power $P_0$. Let the SINR of link $(i,j)$
in $G'(\mathcal{X},P_0,\gamma)$ and $G''(\mathcal{X},P_0,\gamma)$ be
\begin{equation}\label{eq:beta-ij-prime-1}
\beta_{ij}'=\frac{L(d_{ij})}{N_0/P_0+(1+\epsilon_1')\gamma E[J]-\gamma L(d_{ij})}
\end{equation}
and
\begin{equation}\label{eq:beta-ij-prime-prime-1}
\beta_{ij}''=\frac{L(d_{ij})}{N_0/P_0+(1-\epsilon_1)\gamma E[J]-\gamma L(d_{ij})},
\end{equation}
respectively.

Let $C_{ij}'$ and $C_{ij}''$ be the capacity of link $(i,j)$ in $G'(\mathcal{X},P_0,\gamma)$ and $G''(\mathcal{X},P_0,\gamma)$, respectively. Since $\epsilon_1\rightarrow 0$ and $\epsilon_1'\rightarrow 0$ as $n\rightarrow\infty$, $C_{ij}'$ and $C_{ij}''$ are asymptotically equal to $C_{ij}$.

The following lemma establishes a concentration result for $C_k$ with constant transmission power by coupling methods.
\vspace{+.1in}
\begin{lemma}\label{Lemma:C-k-bound-1}
For any $0<\epsilon<1$, the capacity of an $s$-$t$-cut of size $k, k=0,1,...,m$, satisfies
\begin{equation}\label{eq:C-k-concentration-lower-1}
\Pr(C_k\leq (1-\epsilon)E[C_k'])\leq \exp\left\{-\frac{E[C_k']\epsilon^2}{2}\right\}\!\!\left(1\!-\!O\left(\frac{1}{n}\right)\right),
\end{equation}
where $E[C_k']=[m+k(m-k)]\bar{C}'$ and $\bar{C}'$ is the average link capacity in $G'(\mathcal{X},P_0,\gamma)$, and
\begin{equation}\label{eq:C-k-concentration-upper-1}
\Pr(C_k\geq (1+\epsilon)E[C_k''])\leq \exp\left\{-\frac{E[C_k'']\epsilon^2}{3}\right\}\!\!\left(1\!-\!O\left(\frac{1}{n}\right)\right),
\end{equation}
where $E[C_k'']=[m+k(m-k)]\bar{C}''$ and $\bar{C}''$ is the average link capacity in $G''(\mathcal{X},P_0,\gamma)$.
\end{lemma}
\vspace{+.1in}

\emph{Proof:} Since for all $j$, $\{J(j)\geq (1-\epsilon_1)E[J]\}$ and $\{J(j)\leq
(1+\epsilon_1')E[J]\}$ are both increasing events.\footnote{In context of graph theory, an event
$A$ is called increasing if $I_A(G)\leq I_A(G')$ whenever graph $G$ is a subgraph of $G'$, where
$I_A$ is the indicator function of $A$. An event $A$ is called decreasing if $A^{c}$ is
increasing. For details, please see~\cite{Pe03, AlSp00, MeRo96}.} By the FKG inequality
\cite{Pe03, AlSp00, MeRo96}, we have
\begin{small}
\begin{eqnarray*}
\Pr\left(\bigcap_{j=1}^n \{J(j)\geq (1-\epsilon_1)E[J]\}\right)
& \!\!\!\!\!\! \geq & \!\!\!\!\!\! \prod_{j=1}^n\Pr(J(j)\geq (1-\epsilon_1)E[J])\\
&\!\!\!\!\!\! = & \!\!\!\!\!\! \left(1-O\left(\frac{1}{n^2}\right)\right)^n\\
&\!\!\!\!\!\! = & \!\!\!\!\! \! 1-O\left(\frac{1}{n}\right),
\end{eqnarray*}
\end{small}
where the first equality is due to Lemma \ref{Lemma:J-j-bound}, and
\begin{small}
\begin{eqnarray*}
\Pr\left(\bigcap_{j=1}^n \{J(j)\leq (1+\epsilon_1')E[J]\}\right)
& \!\!\!\!\!\!\geq & \!\!\!\!\!\! \prod_{j=1}^n\Pr(J(j)\leq (1+\epsilon_1')E[J])\\
& \!\!\!\!\!\!= & \!\!\!\!\!\! \left(1-O\left(\frac{1}{n^2}\right)\right)^n\\
& \!\!\!\!\!\! = & \!\!\!\!\!\!1-O\left(\frac{1}{n}\right).
\end{eqnarray*}
\end{small}

This implies that $C_{ij}$ is stochastically lower bounded by $C_{ij}'$ and stochastically upper bounded by $C_{ij}''$ with probability $1-O(\frac{1}{n})$. Hence, in order to show \eqref{eq:C-k-concentration-lower-1} and \eqref{eq:C-k-concentration-upper-1}, it suffices to show
\begin{equation}\label{eq:C-k-prime-concentration-1}
\Pr(C_k'\leq (1-\epsilon)E[C_k'])\leq \exp\left\{-\frac{E[C_k']\epsilon^2}{2}\right\}
\end{equation}
and
\begin{equation}\label{eq:C-k-prime-prime-concentration-1}
\Pr(C_k''\geq (1+\epsilon)E[C_k''])\leq \exp\left\{-\frac{E[C_k'']\epsilon^2}{3}\right\}.
\end{equation}

In $G'(\mathcal{X},P_0,\gamma)$ and $G''(\mathcal{X},P_0,\gamma)$, the SINR of link $(i,j)$ is
given by \eqref{eq:beta-ij-prime-1} and \eqref{eq:beta-ij-prime-prime-1}, respectively, and
because $d_{ij}$'s for a given $i$ are independent, by applying the Chernoff bounds, we obtain
\eqref{eq:C-k-prime-concentration-1} and \eqref{eq:C-k-prime-prime-concentration-1}. \qed

Since $C_{ij}'$ and $C_{ij}''$ are asymptotically equal to $C_{ij}$, $E[C_k']$ and $E[C_k'']$ are
asymptotically equal to $E[C_k]$. Consequently, Lemma \ref{Lemma:C-k-bound-1} shows that $C_k$
concentrates at $E[C_k]$ asymptotically almost surely (a.a.s.).

Now, let $C_{s,t}$ be the minimum cut capacity among all $s$-$t$-cuts, i.e., \begin{equation}\label{eq:C-s-t}
C_{s,t}=\min_{0\leq k\leq m}C_k.
\end{equation}
For the given source node $s$ and the sets of destination nodes $\mathcal{T}=\{t_1,...,t_l\}$ and
relay nodes $\mathcal{R}=\{u_1,...,u_m\}$, define the network coding capacity as
\begin{equation}\label{eq:C-s-T}
C_{s,\mathcal{T}}=\min_{t\in
\mathcal{T}}C_{s,t}.
\end{equation}
That is because for one source and multiple destinations, the capacity of network coding depends
on the minimum cut among all the destinations.

In the following, we show that when the number of relay nodes $m$ is sufficiently large, the
network coding capacity $C_{s,\mathcal{T}}$ concentrates at $E[C_0]=m\bar{C}$ with high
probability. \vspace{+.1in}
\begin{theorem}\label{Theorem:Capacity-Lower-Bound-1}
When $n$ is sufficiently large, with high probability, the network coding capacity $C_{s,\mathcal{T}}$ satisfies
\begin{equation}\label{eq:Capacity-Lower-Bound-1}
\Pr(C_{s,\mathcal{T}}\geq (1-\epsilon_{\alpha}')E[C_0])= 1-O\left(\frac{l}{m^{\alpha}}\right),
\end{equation}
where $\epsilon_{\alpha}'=\sqrt{\frac{2\alpha \ln m}{E[C_0]}}$ for $\alpha> 0$ and $E[C_0]=m\bar{C}$.
\end{theorem}
\vspace{+.1in}

\emph{Proof:} Since the $C_{ij}$'s are asymptotically equal to $C_{ij}'$'s, in order to show
\eqref{eq:Capacity-Lower-Bound-1}, it is equivalent to show
\[
\Pr(C_{s,\mathcal{T}}\geq (1-\epsilon_{\alpha}')E[C_0'])= 1-O\left(\frac{l}{m^{\alpha}}\right).
\]

Since $E[C_k']\geq E[C_0']$ for any $k=1,...,m$,
\[
\Pr(C_{s,t}\leq (1-\epsilon_{\alpha}')E[C_0'])\leq \Pr(C_{s,t}\leq
(1-\epsilon_{\alpha}')E[C_{k'}']),
\]
for any $t\in \mathcal{T}$, where $k'$ is the size of the minimum $s$-$t$-cut. By
\eqref{eq:C-k-concentration-lower-1} of Lemma \ref{Lemma:C-k-bound-1}, we have
\begin{small}
\begin{eqnarray*}
\Pr(C_{s,t}\leq (1-\epsilon_{\alpha}')E[C_{k'}']) \!\!\!\!\!\! & \leq & \!\!\!\!\!\!
\exp\left\{-\frac{\epsilon_{\alpha}'^2 [m+k'(m-k')]\bar{C'}}{2}\right\}\\
\!\!\!\!\!\!& \leq & \!\!\!\!\!\!\exp\left\{-\frac{\epsilon_{\alpha}'^2 m\bar{C'}}{2}\right\}.
\end{eqnarray*}
\end{small}
By choosing $\epsilon_{\alpha}'=\sqrt{\frac{2\alpha\ln m}{E[C_0]}}$, since $\bar{C}'$ and $\bar{C}$ are asymptotically equal, we have for any
$t\in \mathcal{T}$,
\[
\Pr(C_{s,t}\leq (1-\epsilon_{\alpha}')E[C_0']) = O\left(\frac{1}{m^\alpha}\right).
\]
By the union bound, we have

\begin{eqnarray*}
\Pr(C_{s,\mathcal{T}}\leq(1-\epsilon_{\alpha}')E[C_0']) \!\!\!\!\!\!& \leq & \!\!\!\!\!\! \sum_{t\in \mathcal{T}}
\Pr(C_{s,t}\leq(1-\epsilon_{\alpha}')E[C_0'])\\
\!\!\!\!\!\! & = & \!\!\!\!\!\! O\left(\frac{l}{m^{\alpha}}\right).
\end{eqnarray*}
\qed

\vspace{+.1in}
\begin{theorem}\label{Theorem:Capacity-Upper-Bound-1}
When $n$ is sufficiently large, with high probability, the network coding capacity $C_{s,\mathcal{T}}$ satisfies
\begin{equation}\label{eq:Capacity-Upper-Bound-1}
\Pr(C_{s,\mathcal{T}}\leq (1+\epsilon_{\alpha}'')E[C_0])= 1-O\left(\frac{1}{m^{\alpha}}\right),
\end{equation}
where $\epsilon_{\alpha}''=\sqrt{\frac{3\alpha \ln m}{E[C_0]}}$ for $\alpha> 0$ and $E[C_0]=m\bar{C}$.
\end{theorem}
\vspace{+.1in}

\emph{Proof:} Since the $C_{ij}$'s are asymptotically equal to $C_{ij}''$'s, in order to show
\eqref{eq:Capacity-Upper-Bound-1}, it is equivalent to show
\[
\Pr(C_{s,\mathcal{T}}\leq (1+\epsilon_{\alpha}'')E[C_0''])= 1-O\left(\frac{1}{m^{\alpha}}\right).
\]
To show this, it is sufficient to consider a particular cut for a pair of the source and one
destination, e.g., an $s$-$t$-cut separating the source $s$ from all the other nodes.
\begin{small}
\begin{eqnarray*}
\Pr(C_{s,\mathcal{T}}\geq(1+\epsilon_{\alpha}'')E[C_0'']) \!\!\!\! & \leq & \!\!\!\! \Pr\left(C_{s,t}\geq(1+\epsilon_{\alpha}'')E[C_0'']\right)\\
\!\!\!\! & \leq & \!\!\!\! \Pr\left(\sum_{i=1}^m
C_{si}\geq(1+\epsilon_{\alpha}'')E[C_0'']\right)\\
\!\!\!\! & = & \!\!\!\! \Pr(C_0 \geq(1+\epsilon_{\alpha}'')E[C_0''])\\
\!\!\!\! & \leq & \!\!\!\! \exp\left(-\frac{\epsilon_{\alpha}''^2 m\bar{C}''}{3}\right)\\
\!\!\!\! & = & \!\!\!\! O\left(\frac{1}{m^{\alpha}}\right).
\end{eqnarray*}
\end{small}
where the last inequality follows from \eqref{eq:C-k-concentration-upper-1} of Lemma
\ref{Lemma:C-k-bound-1}. \qed

\subsection{Heterogeneous Transmission Powers}

In this subsection, we consider the case where the transmission power of each node is random
rather than a constant, but the capacity of a link $(i,j)$ is a constant $R$, which is independent
of the SINR $\beta_{ij}$, when $\beta_{ij}\geq \beta$. In this case, $\beta_{ij}$ can be rewritten
as
\begin{eqnarray}\label{eq:beta-ij-Heterogeneous}
\beta_{ij} & = & \frac{P_iL(d_{ij})}{N_0+\gamma \sum_{k\neq i,j} P_kL(d_{kj})} \nonumber \\
& = & \frac{P_iL(d_{ij})}{N_0+\gamma I(j)-\gamma P_iL(d_{ij})}.
\end{eqnarray}

Because $P_i$'s and $\mathbf{X}_i$'s are both i.i.d., using the same method, we can prove the following lemma:
\vspace{+.1in}
\begin{lemma}\label{Lemma:I-j-bound}
Suppose there are $n$ nodes in the network, then
\begin{equation}\label{eq:I-j-concentration-lower}
\Pr(I(j)\leq (1-\epsilon_2)E[I])=O\left(\frac{1}{n^2}\right),
\end{equation}
and
\begin{equation}\label{eq:I-j-concentration-upper}
\Pr(I(j)\geq (1+\epsilon_2')E[I])=O\left(\frac{1}{n^2}\right),
\end{equation}
for all $j=1,2,...,n$, where $\epsilon_2=\sqrt{\frac{4\ln n}{(n-1)E[P]E[L]}}$ and $\epsilon_2'=\sqrt{\frac{6\ln n}{(n-1)E[P]E[L]}}$.
\end{lemma}

Even though we have concentration results for $I(j)$, we cannot employ the same coupling methods
as in the previous section. This is because in $G'(\mathcal{X},P_0,\gamma)$ (or
$G''(\mathcal{X},P_0,\gamma)$), the $C_{ij}'$'s (respectively, $C_{ij}''$'s) are independent for
all $j\neq i$ for given $i$. In our new case, however, this independence does not hold because all
$C_{ij}$'s depend on transmission power $P_i$. To deal with this dependence, we use martingale
methods and Azuma's inequality to solve our problem.

\vspace{+.1in}
\begin{theorem}[{Azuma's Inequality~\cite{AlSp00}}]\label{Theorem:Azuma's-inequality}
Let $Z_0,Z_1,...,$ be a martingale sequence such that for each $i=1,2,...,$,
\[
|Z_i-Z_{i-1}|\leq c_i
\]
almost surely, where $c_i$ may depend on $i$. Then for all $n>0$ and any $\lambda>0$,

\begin{equation}\label{eq:Azuma's-inequality-lower}
\Pr(Z_n\geq Z_0+\lambda)\leq \exp\left\{-\frac{\lambda^2}{2\sum_{i=1}^nc_i^2}\right\},
\end{equation}
and
\begin{equation}\label{eq:Azuma's-inequality-upper}
\Pr(Z_n\leq Z_0-\lambda)\leq \exp\left\{-\frac{\lambda^2}{2\sum_{i=1}^nc_i^2}\right\}.
\end{equation}
\end{theorem}
\vspace{+.1in}

\emph{Proof:} Please see e.g.~\cite{AlSp00}.\qed

To use Azuma's inequality, we need to construct a martingale. A common approach to obtain a
martingale from a sequence of random variables (not necessarily independent) is to construct a
Doob sequence. More precisely, suppose we have a sequence of random variables $Y_1, Y_2, ...,
Y_n$, which are not necessarily independent. Let $S=\sum_i^n Y_i$ and define a new sequence of
random variables $\{Z_i:i=0,1,...,n\}$ by:
\begin{equation}\label{eq:Doob-sequence}
\left\{\begin{array}{lll}
Z_0 & = & E[S]\\
Z_i & = & E_{Y_{i+1},...,Y_{n}}[S|Y_1,...,Y_i], \quad i=1,2,...,n.
\end{array}\right.
\end{equation}
Then $\{Z_i:i=0,1,...,n\}$ is a martingale and $Z_n=S$.

If we are able to upper bound the difference $|Z_i-Z_{i-1}|$ for all $i$ by some constant, then we
can apply Azuma's inequality to obtain some bound on a tail probability. For example, if $Y_i$'s
are independent, a simple upper bound for $|Z_i-Z_{i-1}|$ is any upper bound on $|Y_i|$. However,
as long as the $Y_i$'s are dependent, which is the case in our model, we cannot bound
$|Z_i-Z_{i-1}|$ in this way. In this case, we need to understand the properties of the $Y_i$'s to
see if we can bound $|Z_i-Z_{i-1}|$. We approach our problem by following this idea and using the
next corollary.

\vspace{+.1in}
\begin{lemma}\label{Lemma:Azuma's-inequality}
For $n>1$, given a sequence of random variables $Y_1, Y_2, ..., Y_n$, which are not necessarily independent, let $S=\sum_i^n Y_i$. If for any $y_i,y_i'\in D_i$, where $D_i$ is the support of $Y_i$,
\[
|E[S|Y_1,...,Y_{i-1},Y_i=y_i]-E[S|Y_1,...,Y_{i-1},Y_i=y_i']|\leq c_i,
\]
almost surely, where $c_i$ may depend on $i$, then for any $\lambda>0$,
\begin{equation}\label{eq:Corollary-inequality-lower}
\Pr(S\geq E[S]+\lambda)\leq \exp\left\{-\frac{\lambda^2}{2\sum_{i=1}^nc_i^2}\right\},
\end{equation}
and
\begin{equation}\label{eq:Corollary-inequality-upper}
\Pr(S\leq E[S]-\lambda)\leq \exp\left\{-\frac{\lambda^2}{2\sum_{i=1}^nc_i^2}\right\}.
\end{equation}
\end{lemma}
\vspace{+.1in}

\emph{Proof:} We prove this corollary for the case of discrete random variables. For continuous
random variables, the proof is similar.

Define a Doob sequence with respect to $\{Y_i:i=1,...,n\}$ as in \eqref{eq:Doob-sequence}. To
simplify the notation, we will write $E_{Y_{i+1},...,Y_{n}}[S|Y_1,...,Y_i]$ as $E[S|Y_1,...,Y_i]$
when there is ambiguity.

By the total conditional probability theorem, we have
\begin{eqnarray*}
Z_{i-1} \!\!\!\!\! & = & \!\!\!\!\! E[S|Y_1,...,Y_{i-1}] \\
\!\!\!\!\! & = & \!\!\!\!\!
\sum_{y\in D_i} E[S|Y_1,...,Y_{i-1},Y_i=y]\Pr(Y_i=y|Y_1,...,Y_{i-1}),
\end{eqnarray*}
and
\begin{eqnarray*}
Z_i  & = &  E[S|Y_1,...,Y_i]\\
 & = & \sum_{y\in D_i} E[S|Y_1,...,Y_i]\Pr(Y_i=y|Y_1,...,Y_{i-1}).
\end{eqnarray*}
Therefore,
\begin{eqnarray*}
& &|Z_i-Z_{i-1}| \\
& = & |E[S|Y_1,...,Y_i]-E[S|Y_1,...,Y_{i-1}]|\\
& = & |\sum_{y\in D_i} E[S|Y_1,...,Y_i]\Pr(Y_i=y|Y_1,...,Y_{i-1}) \\
& & -
\sum_{y\in D_i} E[S|Y_1,...,Y_{i-1},Y_i=y]\Pr(Y_i=y|Y_1,...,Y_{i-1})\\
& \leq & \sum_{y\in D_i} |E[S|Y_1,...,Y_i]-E[S|Y_1,...,Y_{i-1},Y_i=y]|\\
& & \cdot
\Pr(Y_i=y|Y_1,...,Y_{i-1})\\
& \leq & \sum_{y\in D_i} c_i \Pr(Y_i=y|Y_1,...,Y_{i-1})\\
& = & c_i.
\end{eqnarray*}

Since $\{Z_i:i=0,1,...,n\}$ is a martingale with bounded difference of $|Z_i-Z_{i-1}|$, we can
apply Azuma's inequality to obtain \eqref{eq:Corollary-inequality-lower} and
\eqref{eq:Corollary-inequality-upper}. \qed

Now consider $G'(\mathcal{X},\mathcal{P},\gamma)$ and $G''(\mathcal{X},\mathcal{P},\gamma)$
coupled with $G(\mathcal{X},\mathcal{P},\gamma)$ such that they have the same point process
$\mathcal{X}$ and powers $\mathcal{P}$. Then, the SINR of link $(i,j)$ in
$G'(\mathcal{X},\mathcal{P},\gamma)$ and $G''(\mathcal{X},\mathcal{P},\gamma)$ are
\begin{equation}\label{eq:beta-ij-prime-2}
\beta_{ij}'=\frac{P_iL(d_{ij})}{N_0+(1+\epsilon_2')\gamma E[I]-\gamma P_iL(d_{ij})}
\end{equation}
and
\begin{equation}\label{eq:beta-ij-prime-prime-2}
\beta_{ij}''=\frac{P_iL(d_{ij})}{N_0+(1-\epsilon_2)\gamma E[I]-\gamma P_iL(d_{ij})},
\end{equation}
respectively.

Let $C_{ij}'$ and $C_{ij}''$ be the capacity of link $(i,j)$ in $G'(\mathcal{X},\mathcal{P},\gamma)$ and $G''(\mathcal{X},\mathcal{P},\gamma)$, respectively. Then, $C_{ij}'$ and $C_{ij}''$ are asymptotically equal to $C_{ij}$.

Assume that there exist $r_{min}'>0$, $r_{max}'>0$, $r_{min}''>0$ and $r_{max}''>0$ as the solutions for
\[
\frac{p_{min}L(r_{min}')}{N_0+\gamma (1+\epsilon_2')E[I]-\gamma p_{min}L(r_{min}')}=\beta,
\]
\[
\frac{p_{max}L(r_{max}')}{N_0+\gamma (1+\epsilon_2')E[I]-\gamma p_{max}L(r_{max}')}=\beta,
\]
\[
\frac{p_{min}L(r_{min}'')}{N_0+\gamma (1-\epsilon_2)E[I]-\gamma p_{min}L(r_{min}'')}=\beta,
\]
and
\[
\frac{p_{max}L(r_{max}'')}{N_0+\gamma (1-\epsilon_2)E[I]-\gamma p_{max}L(r_{max}'')}=\beta,
\]
respectively. That is
\begin{eqnarray*}
r_{min}' & = & L^{-1}\left(\frac{\beta}{1+\gamma\beta}\cdot\frac{N_0+\gamma(1+\epsilon_2')E[I]}{p_{min}}\right),\\
r_{max}' & = & L^{-1}\left(\frac{\beta}{1+\gamma\beta}\cdot\frac{N_0+\gamma(1+\epsilon_2')E[I]}{p_{max}}\right),\\
r_{min}'' & = & L^{-1}\left(\frac{\beta}{1+\gamma\beta}\cdot\frac{N_0+\gamma(1-\epsilon_2)E[I]}{p_{min}}\right),\\
r_{max}'' & = & L^{-1}\left(\frac{\beta}{1+\gamma\beta}\cdot\frac{N_0+\gamma(1-\epsilon_2)E[I]}{p_{max}}\right).
\end{eqnarray*}

Since $L(\cdot)$ is continuous and strictly decreasing, $r_{min}'$, $r_{max}'$, $r_{min}''$ and
$r_{max}''$ are all unique. In $G'(\mathcal{X},\mathcal{P},\gamma)$
($G''(\mathcal{X},\mathcal{P},\gamma)$), any node inside the circle centered at $\mathbf{X}_i$
with radius $r_{min}'$ ($r_{min}''$) is connected to node $i$ by a bidirectional link; while any
node outside the circle centered at $\mathbf{X}_i$ with radius $r_{max}'$ ($r_{max}''$) is not
connected to node $i$.

Let $\mathcal{A}(\mathbf{X}_i,r_{min}',r_{max}')$ and
$\mathcal{A}(\mathbf{X}_i,r_{min}'',r_{max}'')$ be the two annuli with inner radius $r_{min}'$ and
outer radius $r_{max}'$, and inner radius $r_{min}''$ and outer radius $r_{max}''$, respectively.
Denote by $N(r_{min}',r_{max}')$ and $N(r_{min}'',r_{max}'')$ the number of nodes in
$\mathcal{A}(\mathbf{X}_i,r_{min}',r_{max}')$ and $\mathcal{A}(\mathbf{X}_i,r_{min}'',r_{max}'')$,
respectively. It is clear that $N(r_{min}',r_{max}')$ and $N(r_{min}'',r_{max}'')$ have Poisson
distribution with mean $n\pi(r_{max}'^2-r_{min}'^2)$ and $n\pi(r_{max}''^2-r_{min}''^2)$,
respectively.

Now suppose the signal attenuation function $L(x) = cx^{-\alpha}$ for some constants $c>0$ and
$2<\alpha<4$. Then,
\begin{eqnarray*}
r_{min}' & = & \left(\frac{c(1+\gamma\beta)p_{min}}{\beta[N_0+\gamma(1+\epsilon_2')E[I]]}\right)^{\alpha},\\
r_{max}' & = & \left(\frac{c(1+\gamma\beta)p_{max}}{\beta[N_0+\gamma(1+\epsilon_2')E[I]]}\right)^{\alpha},\\
r_{min}'' & = & \left(\frac{c(1+\gamma\beta)p_{min}}{\beta[N_0+\gamma(1-\epsilon_2)E[I]]}\right)^{\alpha},\\
r_{max}'' & = &
\left(\frac{c(1+\gamma\beta)p_{max}}{\beta[N_0+\gamma(1-\epsilon_2)E[I]]}\right)^{\alpha},
\end{eqnarray*}
and
\begin{eqnarray}
r_{max}'^2-r_{min}'^2 & = &\frac{B(p_{min},p_{max})}{[N_0+\gamma(1+\epsilon_2')E[I]]^{2\alpha}},\label{eq:r-prime-square-difference}\\
r_{max}''^2-r_{min}''^2 & = &
\frac{B(p_{min},p_{max})}{[N_0+\gamma(1-\epsilon_2)E[I]]^{2\alpha}}\label{eq:r-prime-prime-square-difference},
\end{eqnarray}
where
\begin{equation}\label{eq:B}
B(p_{min},p_{max}) =
(p_{max}^{2\alpha}-p_{min}^{2\alpha})\left[\frac{c(1+\gamma\beta)}{\beta}\right]^{2\alpha}.
\end{equation}
>From \eqref{eq:r-prime-square-difference} and \eqref{eq:r-prime-prime-square-difference}, we can
see that both $n\pi(r_{max}'^2-r_{min}'^2)$ and $n\pi(r_{max}''^2-r_{min}''^2)$ scale with $n$ as
$\frac{B}{n^{2\alpha-1}}$, since $E[I]$ scales linear with $n$. Now assume that there exists a
constant $\eta>0$ independent of $n$ such that
\begin{equation}\label{eq:r-max-r-min-constraint}
N(r_{min}',r_{max}') \leq \eta, \quad \mbox{and} \quad N(r_{min}'',r_{max}'')\leq \eta
\end{equation}
hold a.a.s. This assumption actually puts a constraint on the transmission power since it needs to
scale (if it scales) with $n$ so that \eqref{eq:r-max-r-min-constraint} is satisfied. For example,
we may choose $\eta=1$ and the transmission power $P$ scales with $n$ so that
$B(p_{min},p_{max})=\Theta(\frac{n^{2\alpha-1}}{\log n})$. Note that $r_{min}'$ and $r_{max}'$ are
asymptotically equal to $r_{min}''$ and $r_{max}''$, respectively.

The following lemma establishes a concentration result for $C_k$ with heterogeneous transmission
power and constant capacity by coupling methods and Azuma's inequality.

\vspace{+.1in}
\begin{lemma}\label{Lemma:C-k-bound-2}
For any $0<\epsilon<1$, when $n$ is sufficiently large and \eqref{eq:r-max-r-min-constraint} is guaranteed, with high probability, the capacity of an $s$-$t$-cut of size $k, k=0,1,...,m$, satisfies
\begin{equation}\label{eq:C-k-concentration-lower-2}
\Pr(C_k\leq (1-\epsilon)E[C_k'])\leq \exp\left\{-\frac{[m+k(m-k)]\bar{C}'^2\epsilon^2}{2(\eta+1)^2R^2}\right\},
\end{equation}
where $E[C_k']=[m+k(m-k)]\bar{C}'$ and $\bar{C}'$ is the average link capacity in $G'(\mathcal{X},\mathcal{P},\gamma)$, and
\begin{equation}\label{eq:C-k-concentration-upper-2}
\Pr(C_k\geq (1+\epsilon)E[C_k''])\leq \exp\left\{-\frac{[m+k(m-k)]\bar{C}''^2\epsilon^2}{2(\eta+1)^2R^2}\right\},
\end{equation}
where $E[C_k'']=[m+k(m-k)]\bar{C}''$ and $\bar{C}''$ is the average link capacity in $G''(\mathcal{X},\mathcal{P},\gamma)$.
\end{lemma}
\vspace{+.1in}

\emph{Proof:} By Lemma \ref{Lemma:I-j-bound}, for all $j$, $(1-\epsilon_2)E[I]\leq I(j)) \leq (1+\epsilon_2')E[I]$ holds a.a.s. It is clear that $C_{ij}$ is stochastically lower bounded by $C_{ij}'$ and stochastically upper bounded by $C_{ij}''$ almost surely. Hence, in order to show \eqref{eq:C-k-concentration-lower-2} and \eqref{eq:C-k-concentration-upper-2}, it suffices to show
\begin{equation}\label{eq:C-k-prime-concentration-2}
\Pr(C_k'\leq (1-\epsilon)E[C_k'])\leq \exp\left\{-\frac{[m+k(m-k)]\bar{C}'^2\epsilon^2}{2(\eta+1)^2R^2}\right\}
\end{equation}
and
\begin{equation}\label{eq:C-k-prime-prime-concentration-2}
\Pr(C_k''\geq (1+\epsilon)E[C_k''])\leq \exp\left\{-\frac{[m+k(m-k)]\bar{C}''^2\epsilon^2}{2(\eta+1)^2R^2}\right\}.
\end{equation}

To show \eqref{eq:C-k-prime-concentration-2}, we use martingale methods. Let $Y_1=C_{s(k+1)}',
Y_2=C_{s(k+2)}',...,Y_{m-k}=C_{sm}'$, and $Y_{m-k+1}=C_{1(k+1)}', Y_{m-k+2}=C_{1(k+2)}',...,
Y_{m-k+k(m-k)}=C_{km}'$, and $Y_{m-k+k(m-k)+1}=C_{1t}',
Y_{m-k+k(m-k)+2}=C_{2t}',...,Y_{m-k+k(m-k)+k}=C_{kt}'$. Define a Doob sequence $\{Z_i:
i=0,...,m+k(m-k)\}$ with respect to $\{Y_i,i=1,2,...,m+k(m-k)\}$ as
\[\left\{
\begin{array}{lll}
Z_0 & = & E[C_k']\\
Z_i & = & E[C_k'|Y_1,...,Y_i], \quad i=1,2,...,m+k(m-k).
\end{array}\right.
\]
Then $\{Z_i: i=0,...,m+k(m-k)\}$ is a martingale and $Z_{m+k(m-k)}=C_k'$.

Since when $i\neq u$ and $j\neq v$, $C_{ij}'$ is independent of $C_{uv}'$, we have only the
dependence among $C_{ij}'$'s for all $j\neq i$ with given $i$. However, the distance $d_{ij}$'s
are independent for all $j\neq i$ with given $i$. When $d_{ij}\leq r_{min}'$, $C_{ij}'=R$, and
when $d_{ij}> r_{max}'$, $C_{ij}'=0$. Moreover, the number of nodes within the annulus
$\mathcal{A}(\mathbf{X}_i,r_{min}',r_{max}')$ is upper bounded by the constant $\eta$ a.a.s.
Therefore, we have
\[
\begin{array}{rr}
|E[C_k'|Y_1,...,Y_{i-1},Y_i=y_i]-E[C_k'|Y_1,...,Y_{i-1},Y_i=y_i']|\\ \leq
(\eta+1)R
\end{array}
\] a.a.s., where $y_i$ and $y_i'$ are either 0 or $R$. Applying the result of Lemma
\ref{Lemma:Azuma's-inequality}, we have \eqref{eq:C-k-prime-concentration-2}. In the same
manner, we can show that \eqref{eq:C-k-prime-prime-concentration-2} holds. \qed

In the following, we show that as the number of relay nodes $m$ is sufficiently large, the network
coding capacity $C_{s,\mathcal{T}}$ concentrates at $E[C_0]=m\bar{C}$ with high probability. The
proofs are similar to those for Theorem \ref{Theorem:Capacity-Lower-Bound-1} and Theorem
\ref{Theorem:Capacity-Upper-Bound-1}.

\vspace{+.1in}
\begin{theorem}\label{Theorem:Capacity-Lower-Bound-2}
When $n$ is sufficiently large, with high probability, the network coding capacity $C_{s,\mathcal{T}}$ satisfies
\begin{equation}\label{eq:Capacity-Lower-Bound-2}
\Pr(C_{s,\mathcal{T}}\geq (1-\epsilon_{\alpha})E[C_0])= 1-O\left(\frac{l}{m^{\alpha}}\right),
\end{equation}
where $\epsilon_{\alpha}=\frac{(\eta+1)R}{E[C_0]}\sqrt{2\alpha m\ln m}$ for $\alpha> 0$ and $E[C_0]= m\bar{C}$.
\end{theorem}
\vspace{+.1in}

\emph{Proof:} Since $C_{ij}$'s are asymptotically equal to $C_{ij}'$'s, in order to show
\eqref{eq:Capacity-Lower-Bound-2}, it is equivalent to show
\[
\Pr(C_{s,\mathcal{T}}\geq (1-\epsilon_{\alpha})E[C_0'])= 1-O\left(\frac{l}{m^{\alpha}}\right).
\]

Since $E[C_k']\geq E[C_0']$ for any $k=1,...,m$,
\[
\Pr(C_{s,t}\leq (1-\epsilon_{\alpha})E[C_0'])\leq \Pr(C_{s,t}\leq
(1-\epsilon_{\alpha})E[C_{k'}']),
\]
for any $t\in \mathcal{T}$, where $k'$ is the size of the minimum $s$-$t$-cut. By
\eqref{eq:C-k-concentration-lower-2} of Lemma \ref{Lemma:C-k-bound-2}, we have
\begin{small}
\begin{eqnarray*}
\Pr(C_{s,t}\leq (1-\epsilon_{\alpha})E[C_{k'}']) \!\!\!\! & \leq & \!\!\!\!
\exp\left\{-\frac{\epsilon_{\alpha}^2 [m+k'(m-k')]\bar{C'}^2}{2(\eta+1)^2R^2}\right\}\\
\!\!\!\!& \leq & \!\!\!\! \exp\left\{-\frac{\epsilon_{\alpha}^2 m\bar{C'}^2}{2(\eta+1)^2R^2}\right\}.
\end{eqnarray*}
\end{small}
By choosing $\epsilon_{\alpha}=\frac{(\eta+1)R}{E[C_0]}\sqrt{2\alpha m \ln m}$ for $\alpha> 0$, since $\bar{C}'$ and $\bar{C}$ are asymptotically equal, for any
$t\in \mathcal{T}$,
\[
\Pr(C_{s,t}\leq (1-\epsilon_{\alpha})E[C_0']) = O\left(\frac{1}{m^\alpha}\right).
\]
By the union bound, we have
\begin{eqnarray*}
\Pr(C_{s,\mathcal{T}}\leq(1-\epsilon_{\alpha})E[C_0']) \!\!\!\! & \leq & \!\!\!\! \sum_{t\in \mathcal{T}}
\Pr(C_{s,t}\leq(1-\epsilon_{\alpha})E[C_0'])\\
\!\!\!\! & = & \!\!\!\! O\left(\frac{l}{m^{\alpha}}\right).
\end{eqnarray*}\qed

\vspace{+.1in}
\begin{theorem}\label{Theorem:Capacity-Upper-Bound-2}
When $n$ is sufficiently large, with high probability, the network coding capacity $C_{s,\mathcal{T}}$ satisfies
\begin{equation}\label{eq:Capacity-Upper-Bound-2}
\Pr(C_{s,\mathcal{T}}\leq (1+\epsilon_{\alpha})E[C_0])= 1-O\left(\frac{1}{m^{\alpha}}\right),
\end{equation}
where $\epsilon_{\alpha}=\frac{(\eta+1)R}{E[C_0]}\sqrt{2\alpha m \ln m}$ for $\alpha> 0$ and $E[C_0]=m\bar{C}$.
\end{theorem}
\vspace{+.1in}

\emph{Proof:} Since $C_{ij}$'s are asymptotically equal to $C_{ij}''$'s, in order to show
\eqref{eq:Capacity-Upper-Bound-2}, it is equivalent to show
\[
\Pr(C_{s,\mathcal{T}}\leq (1+\epsilon_{\alpha})E[C_0''])= 1-O\left(\frac{1}{m^{\alpha}}\right).
\]
To show this, it is sufficient to consider a particular cut for a pair of the source and one
destination, for instance, an $s$-$t$-cut separating the source $s$ from all the other nodes.
\begin{small}
\begin{eqnarray*}
\Pr(C_{s,\mathcal{T}}\geq(1+\epsilon_{\alpha})E[C_0'']) \!\!\!\! & \leq & \!\!\!\! \Pr\left(C_{s,t}\geq(1+\epsilon_{\alpha})E[C_0'']\right)\\
\!\!\!\! & \leq & \!\!\!\! \Pr\left(\sum_{i=1}^m
C_{si}\geq(1+\epsilon_{\alpha})E[C_0'']\right)\\
\!\!\!\! & = & \!\!\!\! \Pr(C_0 \geq(1+\epsilon_{\alpha})E[C_0''])\\
\!\!\!\! & \leq & \!\!\!\! \exp\left\{-\frac{\epsilon_{\alpha}^2 m\bar{C}''^2}{2(\eta+1)^2R^2}\right\}\\
\!\!\!\! & = & \!\!\!\! O\left(\frac{1}{m^{\alpha}}\right).
\end{eqnarray*}
\end{small}
where the last inequality follows from \eqref{eq:C-k-concentration-upper-2} of Lemma
\ref{Lemma:C-k-bound-2}.\qed

\section{Network Coding Capacity for Multiple-Source Transmission}

In this section, we study network coding capacity for multiple sources and multiple destinations
transmission. We assume the same notation as in Section III. However, instead of having a single
source, we have $s\geq 2$ sources. Denote by $\mathcal{S}=\{s_1,\dots,s_h\}$ the set of source
nodes. Assume there is no correlation among the set of sources $\mathcal{S}$. Now we can define an
$\mathcal{S}$-$t$-cut of size $k$  between the set of sources $\mathcal{S}$ and one destination
$t\in \mathcal{T}$ as a partition of the relay nodes into two sets $V_k$ and $V_k^c$, such that
$|V_k|=k,|V_k^c|=m-k$, $V_k \cup V_k^c=\mathcal{R}$ and $V_k\cap V_k^c=\emptyset$. Let
\begin{equation}\label{eq:C-k-multiple}
C_k=\sum_{i=1}^h \sum_{u_j\in V_k^c}C_{s_ij}+\sum_{u_j\in V_k}\sum_{u_i\in
V_k^c}C_{ji}+\sum_{u_j\in V_k}C_{jt},
\end{equation}
then $C_k$ is the capacity of the corresponding $\mathcal{S}$-$t$-cut, and
\begin{eqnarray}\label{eq:C-k-expectation-multiple}
E[C_k] & = & E_{X}E_P[C_k]\nonumber\\
& = & [(m-k)h+k(m-k)+k]\bar{C},
\end{eqnarray}

Now, let $C_{\mathcal{S},t}$ be the minimum cut capacity among all $\mathcal{S}$-$t$-cuts, \begin{equation}\label{eq:C-S-t}
C_{\mathcal{S},t}=\min_{0\leq k\leq m}C_k.
\end{equation}
By comparing \eqref{eq:C-k-expectation} and \eqref{eq:C-k-expectation-multiple}, we note that we
on longer have symmetry in $E[C_k]$ with respect to $k$, i.e., $E[C_k]\neq E[C_{m-k}]$ for
$k=0,1,...,m$. In the single source case, the minimum value of $E[C_k]$, i.e., $E[C_{s,t}]$ is
obtained when $k=0$ or $k=m$ due to the symmetry ($E[C_0]=E[C_m]$). This means that the
bottlenecks are at the source end and also the destination end. Nevertheless, when we have
multiple sources, $E[C_0]>E[C_m]$, and the minimum expectation value of the capacity among all
cuts with any size is $E[C_{\mathcal{S},t}]=E[C_m]$, which implies that we have only one
bottleneck at the destination end.

For the given set of source nodes $\mathcal{S}=\{s_1,...,s_s\}$ and the sets of destination nodes
$\mathcal{T}=\{t_1,...,t_l\}$ and relay nodes $\mathcal{R}=\{u_1,...,u_m\}$, define the network
coding capacity for multiple sources and multiple destinations as
\begin{equation}\label{eq:C-S-T}
C_{\mathcal{S},\mathcal{T}}=\min_{t\in \mathcal{T}}C_{\mathcal{S},t}.
\end{equation}
Then, by the same method used in the previous section, we can show that
$C_{\mathcal{S},\mathcal{T}}$ concentrates at $E[C_m]=m\bar{C}$ with high probability, where
$\bar{C}$ is defined the same as before. This indicates that $C_{\mathcal{S},\mathcal{T}}$ and
$C_{s,\mathcal{T}}$ concentrate at the same value. This is because they have one bottleneck in
common.

\vspace{+.1in}
\begin{theorem}\label{Theorem:Capacity-Lower-Bound-3}
When $n$ is sufficiently large, the network coding capacity $C_{\mathcal{S},\mathcal{T}}$ satisfies
\begin{equation}\label{eq:Capacity-Lower-Bound-3}
\Pr(C_{\mathcal{S},\mathcal{T}}\geq (1-\epsilon_{\alpha})E[C_m])= 1-O\left(\frac{l}{m^{\alpha}}\right),
\end{equation}
where $\epsilon_{\alpha}=\frac{(\eta+1)R}{E[C_m]}\sqrt{2\alpha m\ln m}$ for $\alpha> 0$ and $E[C_m]= m\bar{C}$.
\end{theorem}
\vspace{+.1in}

\emph{Proof:} The proof is the same as that for Theorem \ref{Theorem:Capacity-Lower-Bound-2} by
replacing of $E[C_0]$ by $E[C_m]$\qed

\vspace{+.1in}
\begin{theorem}\label{Theorem:Capacity-Upper-Bound-3}
When $n$ is sufficiently large, the network coding capacity $C_{\mathcal{S},\mathcal{T}}$ satisfies
\begin{equation}\label{eq:Capacity-Upper-Bound-3}
\Pr(C_{\mathcal{S},\mathcal{\mathcal{T}}}\leq (1+\epsilon_{\alpha})E[C_m])= 1-O\left(\frac{1}{m^{\alpha}}\right),
\end{equation}
where $\epsilon_{\alpha}=\frac{(\eta+1)R}{E[C_m]}\sqrt{2\alpha m\ln m}$ for $\alpha> 0$ and $E[C_m]= m\bar{C}$.
\end{theorem}
\vspace{+.1in}

\emph{Proof:} The proof is the same as that for Theorem \ref{Theorem:Capacity-Upper-Bound-2} by
replacing of $E[C_0]$ by $E[C_m]$\qed

\begin{figure}[t!]
\centering
\includegraphics[width=2.6in]{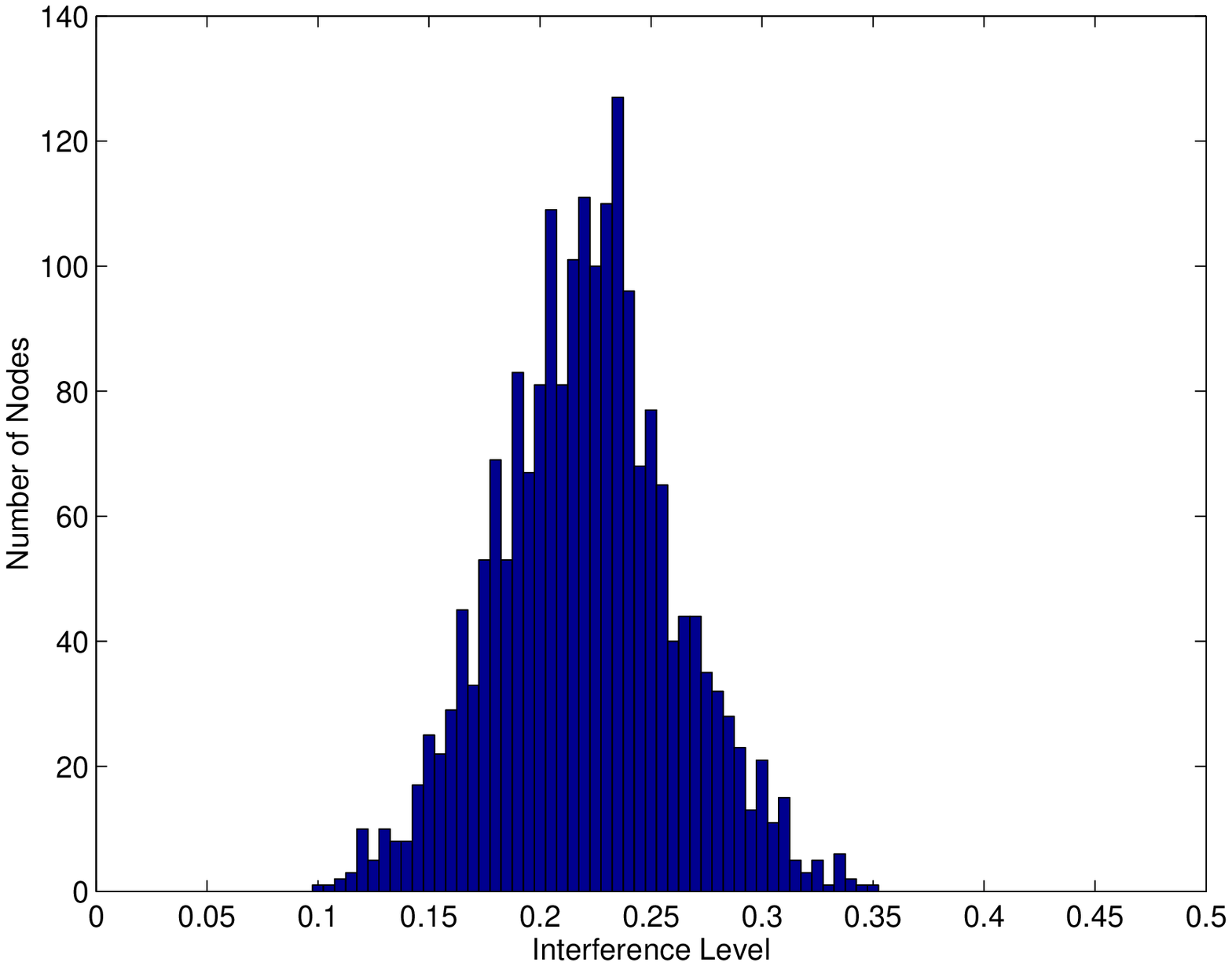}
\caption{Interference at each node in $G(\mathcal{X},P_0,\gamma)$}
\end{figure}

\begin{figure}[t!]
\centering
\includegraphics[width=2.6in]{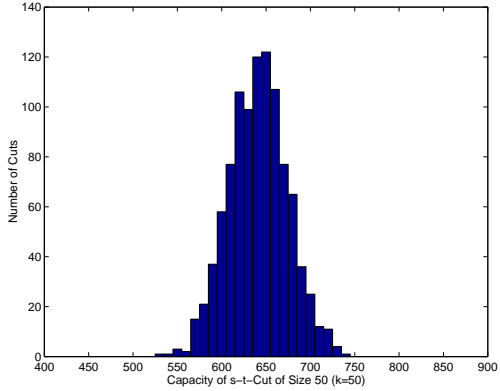}
\caption{Capacity of random s-t-cut of size $k=50$ in $G(\mathcal{X},P_0,\gamma)$}
\end{figure}

\section{Simulation Studies}

In this section, we present some simulation results on the SINR model and network coding capacity.
Fig.~3 and Fig.~4 show simulation results on interference and cut capacity in
$G(\mathcal{X},P_0,\gamma)$, where $n=2000$, $L(x)=\frac{10^{-3}}{64}x^{-3}$, $N_0 = 0.02$, $\beta
= 0.2$ and $\gamma = 0.02$, and every node transmits with constant power $P_0=0.01$. Fig.~5 and
Fig.~6 show simulation results on interference and cut capacity in
$G(\mathcal{X},\mathcal{P},\gamma)$, where $n=2000$, $L(x)=\frac{10^{-3}}{64}x^{-3}$, $N_0 =
0.02$, $\beta = 0.2$ and $\gamma = 0.02$, and every node transmits with power $P$ uniformly
randomly distributed over $[0.01,0.02]$. The results confirm the concentration behavior of
interference and cut capacity.

\begin{figure}[t]
\centering
\includegraphics[width=2.6in]{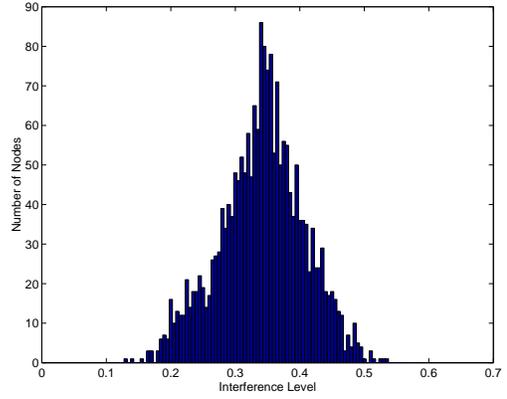}
\caption{Interference at each node in $G(\mathcal{X},\mathcal{P},\gamma)$}
\end{figure}

\begin{figure}[t]
\centering
\includegraphics[width=2.6in]{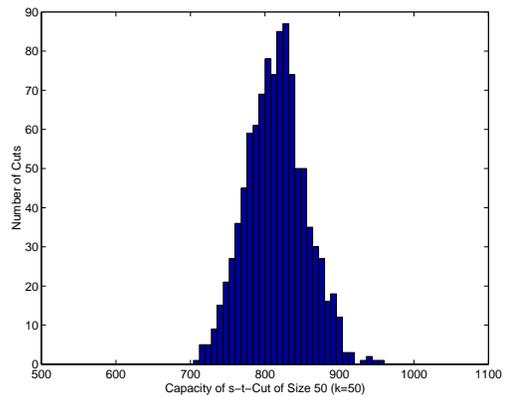}
\caption{Capacity of random s-t-cut of size $k=50$ in $G(\mathcal{X},\mathcal{P},\gamma)$}
\end{figure}

\section{Conclusions}

In this paper, we studied network coding capacity for random wireless
networks with interference and noise. In this model, the capacities
of links are not independent. By using coupling and martingale
methods, we showed that when the size of the network is sufficiently
large, the network coding capacity still exhibits a concentration
behavior in cases of single source multiple destinations and multiple
sources multiple destinations. We demonstrated simulation results
that meet our theoretical bounds of network coding capacity.

\end{document}